\documentclass[]{piparticle-final}
\usepackage{graphicx}
\usepackage{amsmath}
\usepackage{cite} 
\usepackage{epstopdf} 

\begin{document}

\title{High-speed tunable photonic crystal fiber-based femtosecond soliton source without dispersion pre-compensation}
\author{Mart\'\i{}n Caldarola,\cite{lec}\thanks{E-mail: caldarola@df.uba.ar}\hspace{.5em}  
        V\'\i{}ctor A. Bettachini,\cite{itba} 
        Andr\'es A. Rieznik,\cite{itba}
	Pablo G. K\"onig,\cite{itba}
	Mart\'\i{}n E. Masip,\cite{lec}
	Diego F. Grosz,\cite{itba,coni}
	Andrea V. Bragas\cite{lec, ifiba}
	}

\volume{4}               
\articlenumber{040001}   
\journalyear{2012}       
\editor{A. Go\~{n}i}   
\reviewers{J. Chavez Boggio, Leibniz Institut f\"{u}r Astrophysik Potsdam, Germany.}  
\received{7 July 2011}     
\accepted{1 February 2012}   
\runningauthor{M Caldarola \itshape{et al.}}  
\doi{040001}         

\pipabstract{
We present a high-speed wavelength tunable photonic crystal fiber-based source capable of generating tunable femtosecond solitons in the infrared region.
Through measurements and numerical simulation, we show that both the pulsewidth and the spectral width of the output pulses remain nearly constant over the entire tuning range from $860$ to $1160$~nm.
This remarkable behavior is observed even when pump pulses are heavily chirped ($7400$~fs$^2$), which allows to avoid bulky compensation optics, or the use of another fiber, for dispersion compensation usually required by the tuning device.
}

\maketitle

\blfootnote{
\begin{theaffiliation}{99}
   \institution{lec} Laboratorio de Electr\'onica Cu\'antica, Departamento de F\'\i{}sica, Universidad de Buenos Aires, Pabell\'on I, Ciudad Universitaria, (C1428EHA) Buenos Aires, Argentina.
   \institution{itba} Instituto Tecnol\'ogico de Buenos Aires, Eduardo Madero 399, (C1106ACD) Buenos Aires, Argentina.
   \institution{coni} Consejo Nacional de Investigaciones Cient\'\i{}ficas y T\'ecnicas, Argentina.
   \institution{ifiba} IFIBA, Consejo Nacional de Investigaciones Cient\'\i{}ficas y T\'ecnicas, Argentina.

\end{theaffiliation}
}

\section{Introduction}\label{sec:intro}

Light sources based on the propagation of solitons in optical fibers have emerged as a compact solution to the need of a benchtop source of ultra-short tunable pulses \cite{Nishizawa1999,Abedin2004a,Lee2008}. 
The soliton formation from femtosecond pulses launched into an optical fiber is explained in terms of the interplay between self-phase modulation (SPM) and group-velocity dispersion (GVD) in the anomalous dispersion regime \cite{Agrawal}.
The wavelength tunability is a consequence of the Raman-induced frequency shift (RIFS) produced on the pulse when traveling through the fiber \cite{Mitschke:86}.
The term soliton self-frequency shift (SSFS) \cite{soliton-teo-Gordon} was coined to name this effect widely used to produce tunable femtosecond pulses in different wavelength ranges, e.g., from $850$ to $1050$~nm \cite{Washburn2001}, from $1050$ to $1690$~nm \cite{Takayanagi:06}, and from $1566$ to $1775$~nm \cite{Nishizawa1999}.
In most cases, photonic crystal fibers (PCF) are used for building these sources since their GVD can be easily tailored to produce solitons in a desired tuning range \cite{Russell03, Skryabin03}. 
For a given choice of the PCF, full experimental characterization of the pump and output pulses, complemented with theoretical predictions, is necessary to understand how nonlinear effects modify the output soliton.

The wavelength tunability in a PCF-based light source is provided by the modulation of the pump power injected into the fiber \cite{Nishi02,Abedin:03,Ishii2006,fs_soliton_source}. 
It is worth noting that the wavelength choice of the output pulse is done without moving any mechanical part, which is clearly attractive for all the proposed and imaginable applications of these soliton sources. 
Moreover, the wavelength of the output pulse can be chosen as fast as one can modulate the power of the pump pulse, as introduced in Ref. \cite{Sanders2002,Walewski2004}.
By introducing an acousto-optic modulator (AOM) in the path of the pump pulse, the output wavelength can be changed at a speed which is ultimately limited only by the laser repetition rate. 
This kind of experimental setup has been presented in some previous reports \cite{Sumimura:08,fs_soliton_source}, with stunning applications as the one presented in Ref. \cite{Sumimura2010}, where a pseudo-CW wideband source for optical coherent tomography is introduced.
However, the need to pre-compress the pump pulse to avoid the chirp produced by the AOM contrives against the compact and mechanically robust design of the light source. 
In this paper, we demonstrate that the PCF-based source presented here is robust against chirped pump pulses.
We present a complete set of measurements showing that the temporal and spectral characteristics of the generated solitons in the PCF remain unaltered even when pump pulses are heavily chirped up to $\sim7400$~fs$^2$. Results are presented for the whole range of tunability ($860$~nm to $1160$~nm).
We also present numerical simulations which remarkably fit the experimental data and help to understand the soliton behavior. 

This paper is organized as follows: In section II, we describe the experimental setup. The numerical simulations are described in section III.
In section IV, we present experimental and numerical results and in section V we further analyze the results with numerical simulations.
Finally, in section VI, we present our conclusions.

\section{Experimental Setup}\label{sec:exp}

\begin{figure}[htbp] 
  \centering \includegraphics[width=0.45\textwidth]{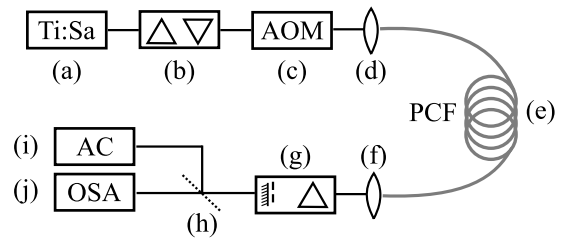}
  \caption{Experimental setup.
    (a) Titanium-Sapphire (Ti:Sa) laser,
    (b) Prism compressor,
    (c) Acousto-optic modulator (AOM),
    (d) Coupling lens,
    (e) Photonic crystal fiber (PCF),
    (f) Collimator objective,
    (g) Spatial filter,
    (h) Flipper mirror,
    (i) Fast-scan interferometric autocorrelator,
    (j) Optical spectrum analyzer (OSA).
  }
  \label{fg:setup}
\end{figure}

A scheme of the experimental setup is shown in Fig. \ref{fg:setup}. 
A Ti:Sa laser (KMLabs) generates ultrashort transform-limited (TL) pulses of $\mathrm{\Delta t}=31$~fs (FWHM-sech$^2$), $\mathrm{\lambda_{pump}}=830$~nm, with a spectral width $\mathrm{\Delta \lambda}= 23$~nm, and a repetition rate of $94$~MHz.

The AOM not only allows high speed (up to MHz) and accurate control of the soliton wavelength, as previously discussed, but also prevents feedback into the Ti:Sa, replacing the optical isolator required in similar setups \cite{chan:08}.
As the AOM introduces $\sim56$~mm of SF8 glass path, pump pulses gain a positive chirp of about $\sim7400$~fs$^2$, which leads to a time spread by a factor of $\sim3$ in them. 
This can be pre-compensated, for example, by introducing an optical fiber in the anomalous dispersion regime \cite{Nicholson:04,Takayanagi:06} or a prism compressor in the well-known configuration presented in \cite{prisma_OEM}.
In this work, the chirp was compensated by a pair of SF18 prisms with an apex separation of $78$~cm.
Additionally, the prism compressor allowed us to up-chirp pump pulses in a controlled fashion from TL to $\sim1400$~fs$^2$ by introducing an extra glass path at the second prism of the arrangement \cite{brito-cruz}. 
This full or partial compensation of the phase distortion introduced by the AOM allowed us to study the role of different chirp figures in the temporal and spectral characteristics of the solitons generated in the PCF.

\begin{figure}[htbp] 
   \centering \includegraphics[width=0.45\textwidth]{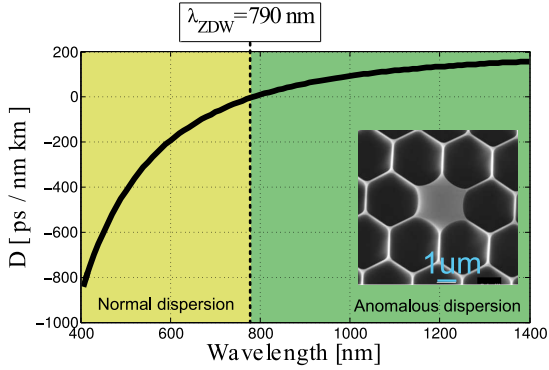}
   \caption{Dispersion curve of the PCF, showing the zero dispersion wavelength (ZDW) at $790$~nm. The inset is the scanning electron microscope image of the PCF core. The curve and image were provided by the manufacturer.}
   \label{fg:beta}
 \end{figure}

Pump pulses were coupled into a non-polarization-maintaining microstructured fiber commercially used for supercontinuum generation (Thorlabs, NL-2.3-790-02). Its main parameters are listed in Table \ref{tb:pcf} and the dispersion curve and SEM image are shown in Fig. \ref{fg:beta}.\footnote{Datasheet available in http://www.thorlabs.com.}
Upon propagation down the fiber, the spectrum is highly broadened so a spatial band-pass filter made of a prism and razor blades, similar to the one presented in \cite{Chilla:91}, allowed to filter the spectral region of the solitonic branch (see Fig. \ref{fg:spec_P_con}) without adding any extra chirp to the solitons.

\begin{table}[tbp]
  \centering
  \begin{tabular}{|l|c|}
    \hline
    L & $75$~cm\\
    ZDW & $790$~nm \\
    $\beta_2$ & $-12.4$~ps$^2$km$^{-1}$ \\
    $\beta_3$  & $0.07$~ps$^3$km$^{-1}$ \\
    $\gamma(\omega)$ & $\gamma_0(\omega_0)+ (\omega- \omega_0) \gamma_1$\\
    $\gamma_0(\omega_0)$ & $78$~W$^{-1}$km$^{-1}$ \\
    $\gamma_1$ & $\gamma_0/ \omega_0$ \\
    $\omega_0$ & $2271$~THz \\
    \hline
    \end{tabular}
  \caption{PCF parameters relevant to the simulation. Further details can be found in Ref. \cite{fs_soliton_source}.}\label{tb:pcf}
\end{table}

Once the spectral selection was achieved, a flipper mirror directed the filtered beam for analysis either by the optical spectrum analyzer (OSA) or by the interferometric autocorrelator.
A fast-scan system \cite{Costantino_fs} allows to perform fast interferometric auto-correlations. 
Briefly, a platform with a hollow retroreflector is moved sinusoidally back and forth, with a stepper motor at 11 Hz, to produce and optical delay in one of the arms of a Michelson interferometer. 
The autocorrelation signal is recorded by a PMT and averaged with an oscilloscope.

\section{Numerical Simulations} \label{sec:sim}
In order to further validate experimental results, we simulated the propagation of femtosecond pulses in the PCF by numerically solving the generalized nonlinear Schr\"odinger equation (GNLSE) including dispersive, Kerr, instantaneous and delayed Raman response, and self-steepening effects \cite{Dudley_06}, with a conservation quantity error (CQE) adaptive step-size algorithm \cite{Heidt:09}.

The GNLSE reads 

\begin{align}
  &\frac{\partial A}{\partial z} + \beta_1 \frac{\partial A}{\partial t} + \operatorname{i} \beta_2 \frac{\partial^2 A}{\partial t^2} \label{eq:gnlse} \\ 
  & - \beta_3 \frac{\partial^3 A}{\partial t^3}+ ... = \operatorname{i} \gamma(\omega) \left(1+ \frac{\operatorname{i}}{\omega_0} \frac{\partial}{\partial t} \right)  \notag\\
  & \times \left( A(z,t) \int_{-\infty}^\infty R(t') |A(z,t- t')|^2 dt' \right), \notag\\
  \notag \\
  & R(t)= \left(1- f_R \right) \delta(t)+ f_R h_R (t),\notag\\
  & h_R(t)= \left(f_a+ f_c \right) h_a(t) + f_b h_b(t),\notag\\
  & h_a(t)= \tau_1 \left( \tau_1^{-2}+ \tau_2^{-2} \right) \operatorname{e}^{-t/\tau_2} \sin{(t/\tau_1)}, \notag\\
  & h_b(t)= \left[ \left(2\tau_b- t\right)/ \tau_b^2\right] \operatorname{e}^{-t/\tau_b}, \notag
  \end{align}
where $A(z,t)$ is the complex envelope of the electric field, $\beta_n$ are the expansion terms for the propagation constant around the carrier frequency $\omega_0$ and $\gamma$ is the nonlinear coefficient.
  $f_R(t)$ represents the fractional contribution of the delayed Raman effect $h_R$.
Note that Eq. (\ref{eq:gnlse}) adopts a more accurate description of this effect than the one usually used \cite{Agrawal}.
In our simulation, we adopted $\tau_1= 12.2$~fs, $\tau_2= 32$~fs, $\tau_b= 96$~fs, $f_a= 0.75$, $f_b= 0.21$, $f_c= 0.04$, and $f_R= 0.24$ \cite{Lin_06}.
The dependence of the fiber non-linear parameter $\gamma$ with the frequency was modeled as a linear function (see Table \ref{tb:pcf}).

\section{RESULTS} \label{sec:res}

\subsection{Transform-limited pump pulses}\label{sec:TL}

First, we present the full characterization of the soliton source seeded by TL pump pulses, in an extended wavelength range if compared with the results presented in our previous paper \cite{fs_soliton_source}.

\begin{figure}[htbp]
  \includegraphics[width=0.45\textwidth]{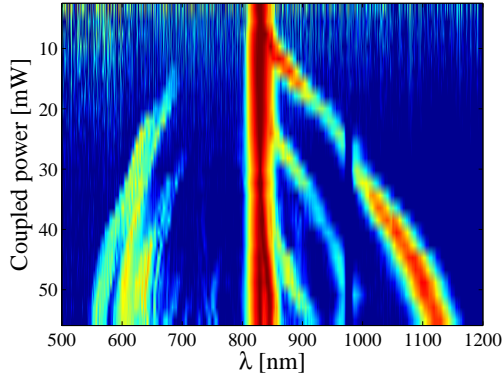} 
  \caption{Experimental spectra vs coupled power to the PCF with transform limited (TL) pump pulses. The color map shows spectral intensity. The maximum achieved soliton shift, $\lambda_s \simeq1130$~nm, was reached at $55$~W.}
  \label{fg:spec_P_con}
\end{figure}

In order to investigate the dependence of the output spectrum with the coupled power, managed by the AOM, we skipped spectral filtering at first.
Fig. \ref{fg:spec_P_con} shows the measured spectrum at the PCF output as a function of the coupled power.
The infrared solitonic branch appears at $\sim10$~mW and undergoes red-shift with increasing power.
The maximum wavelength attained is $1130$~nm at $55$~mW.
Spectra in Fig. \ref{fg:spec_P_con} also shows show that some of the input energy is converted to visible non-solitonic radiation.

\begin{figure}[htbp]
  \includegraphics[width= 0.45\textwidth]{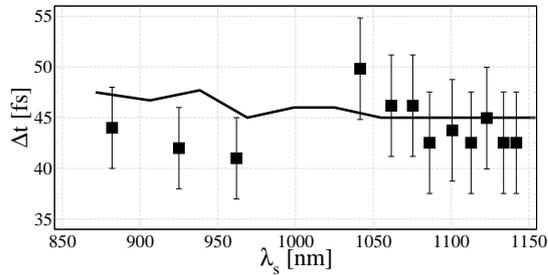}
  \caption{Experimental pulsewidth of the soliton as a function of its wavelength, pumping the PCF with TL pulses. The results for the three lower wavelengths were already present in Ref. \cite{fs_soliton_source}. Full line: numerical simulations.}
  \label{fg:dt_vs_lo_TL}
\end{figure}

The pulsewidth of the filtered soliton as a function of its wavelength, $\lambda_s$, is shown in Fig. \ref{fg:dt_vs_lo_TL}.
The pulsewidth remains constant at $\sim45$~fs, for the entire tunability range.
Numerical simulations are also plotted in the same figure, showing an excellent agreement with experimental measurements. 

\subsection{Chirped pump pulses} \label{sec:chirp}

The effect over the soliton produced by the chirp of pump pulses was studied systematically by introducing a known amount of extra glass path on the second prism of the compressor.
 This scheme allowed to change the GVD of pump pulses from $0$ to $1400$~fs$^2$. 
Further chirping was achieved by the complete removal of the prism compressor, leading to a total amount of positive chirp $\sim 7400$~ fs$^2$.

\begin{figure}[htbp]
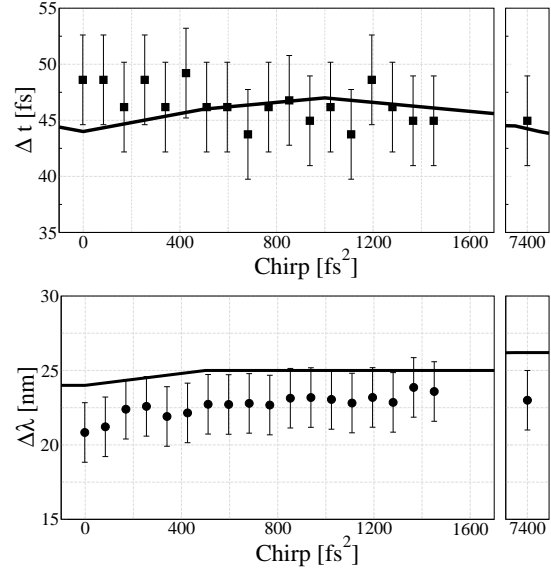

  \includegraphics[width= 0.45\textwidth]{Caldarola_fig5}
  \includegraphics[width= 0.45\textwidth]{Caldarola_fig6}
  \caption{ Soliton temporal (a) and spectral width (b) vs. chirp of input pump pulses . Full line: numerical results. The soliton wavelength is $\lambda_s=1075$~nm.}
  \label{fg:chirp_dt_dl}
\end{figure}

Figure \ref{fg:chirp_dt_dl} (a) shows the pulsewidth of solitons with wavelength $\lambda_s=1075$~nm upon variation of pump pulses chirp.
Even for a $\sim7400$~fs$^2$ chirp, the soliton output pulsewidth remained around $45$~fs.
Numerical simulations show very good agreement with these observations, as they predict nearly constant pulsewidth regardless of the input chirp (full line in Fig. \ref{fg:chirp_dt_dl}). 
Measurements and numerical simulations in the spectral domain also indicate that the bandwidth of the output solitons is almost unaffected by the pump pulses chirp [see Fig. \ref{fg:chirp_dt_dl} (b)].
The product $\Delta t \Delta\nu$ was found to be near $0.315$, as it is expected for transform-limited sech$^2$ pulses.

\begin{figure}[htbp]
  \includegraphics[width=0.45\textwidth]{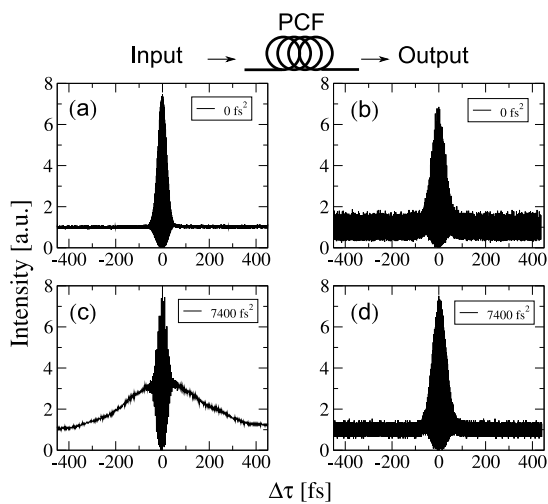}
  \caption{
    Interferometric auto-correlation traces of TL (a) and heavily chirped, $\sim7400$~fs$^2$, Ti:Sa pump pulses (c).
    Interferometric auto-correlation traces of the output solitons are similar in both cases, unchirped (b) and heavily chirped (d).
    Soliton wavelength is $\lambda_s\simeq1075$~nm.}
  \label{fg:chirpac}
\end{figure}

The effect of this heavy chirping was evident in the auto-correlation traces of pump pulses, as can be seen by comparing Fig. \ref{fg:chirpac} (a) and (c).
However, there is not a clear difference between traces of the output solitons for the TL (b) and the highly chirped ($\sim7400$~fs$^2$) case (d).

\begin{figure}[htbp]
  \includegraphics[width=0.45\textwidth]{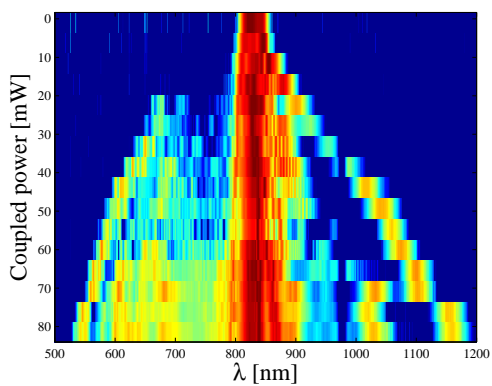} 
  \caption{Spectra vs coupled power to the PCF with highly chirped ($\sim7400$~fs$^2$) input pulses.}
  \label{fg:spec_P_sin}
\end{figure}

A color map of the spectra as function of the coupled power, for a highly chirped pump pulse ($\sim7400$~fs$^2$), is shown in Fig. \ref{fg:spec_P_sin}. 
As in the TL case, we observe that a solitonic branch is red shifted by increasing the coupled power. 
However, in this case, $80$~mW of coupled power is required to produce a $1160$~nm soliton which represents an increment of about $\sim 45\%$ in comparison to the TL case.

\begin{figure}[htbp]
  \centering
  \includegraphics[width=0.45\textwidth]{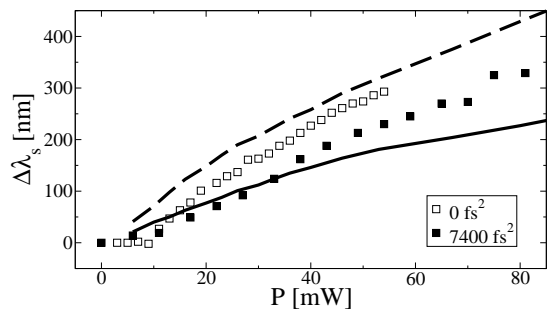}
  \caption{Soliton wavelength shift for chirped (full squares) and TL pump pulses (empty squares) vs pump pulse power.
    Dashed and full lines correspond to numerical results for TL and highly chirped pump pulses, respectively.
    }
  \label{fg:lo_pw} 
\end{figure}

A comparison of the soliton red-shift between TL and chirped pump pulse cases is presented Fig. \ref{fg:lo_pw}.
The figure shows that more power is always required to attain the same shift when pump pulses are heavily chirped.

\begin{figure}[htbp]
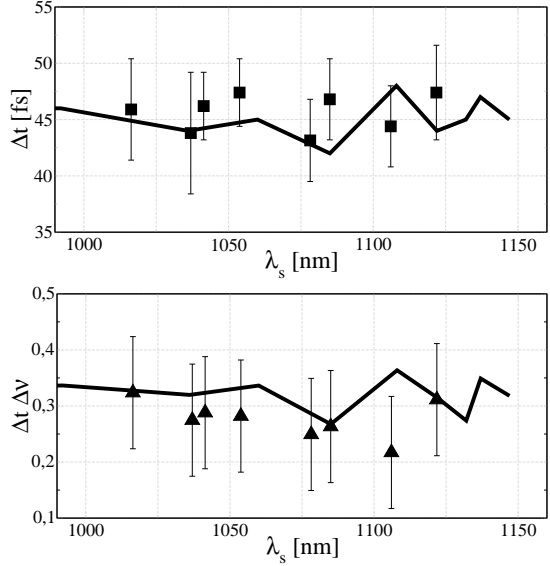

  \includegraphics[width=0.45\textwidth]{Caldarola_fig10}
  \includegraphics[width=0.45\textwidth]{Caldarola_fig11}
  \caption{(a) Soliton pulsewidth and (b) the product $\mathrm{\Delta t \Delta \nu}$ vs wavelength in the case of highly chirped pump pulses ($\sim$7400~fs$^2$).}
  \label{fg:Dt_vs_loWO}
\end{figure}

Figure \ref{fg:Dt_vs_loWO} (a) shows the soliton pulsewidth as a function of its wavelength, $\lambda_s$, when the pump pulse is heavily chirped ($\sim7400$~fs$^2$).
We observe an approximately constant output pulsewidth ($\sim45$~fs) in the entire tuning range.
Furthermore, the $\mathrm{\Delta t \Delta \nu}$ product, shown in Fig. \ref{fg:Dt_vs_loWO} (b), indicates that the generated pulses can be identified as fundamental solitons (sech$^2$-like), as in the case of TL pump pulses \cite{fs_soliton_source}.
Numerical simulations were also performed for this case (full lines in Fig. \ref{fg:Dt_vs_loWO}) showing an excellent agreement with experimental results. 

\section{DISCUSSION}\label{sec:disc}

\subsection{Fiber soliton self-frequency shift effective length} \label{sec:len} 
\begin{figure}[htbp]
  \centering
  \includegraphics[width=0.45\textwidth]{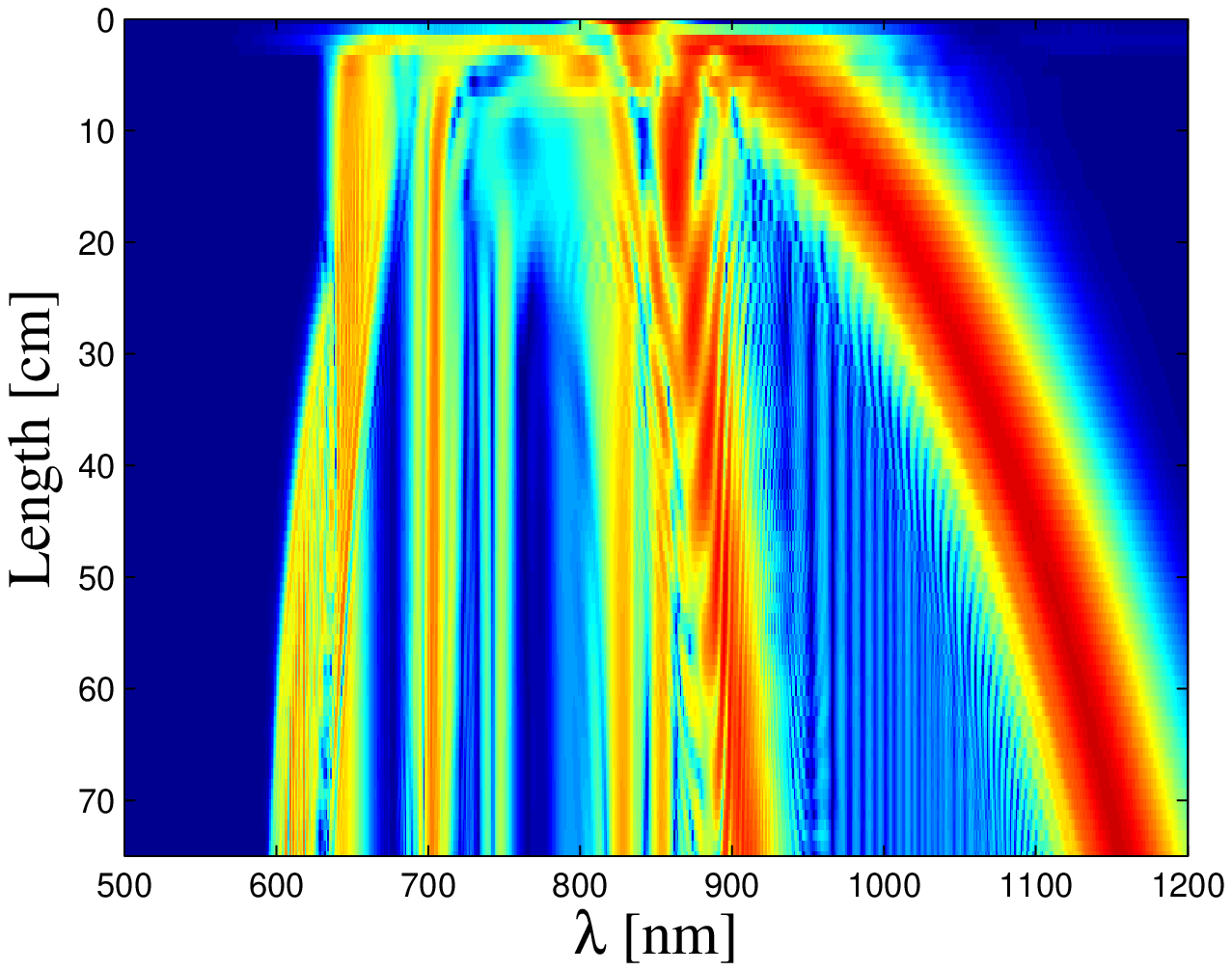}
  \includegraphics[width=0.45\textwidth]{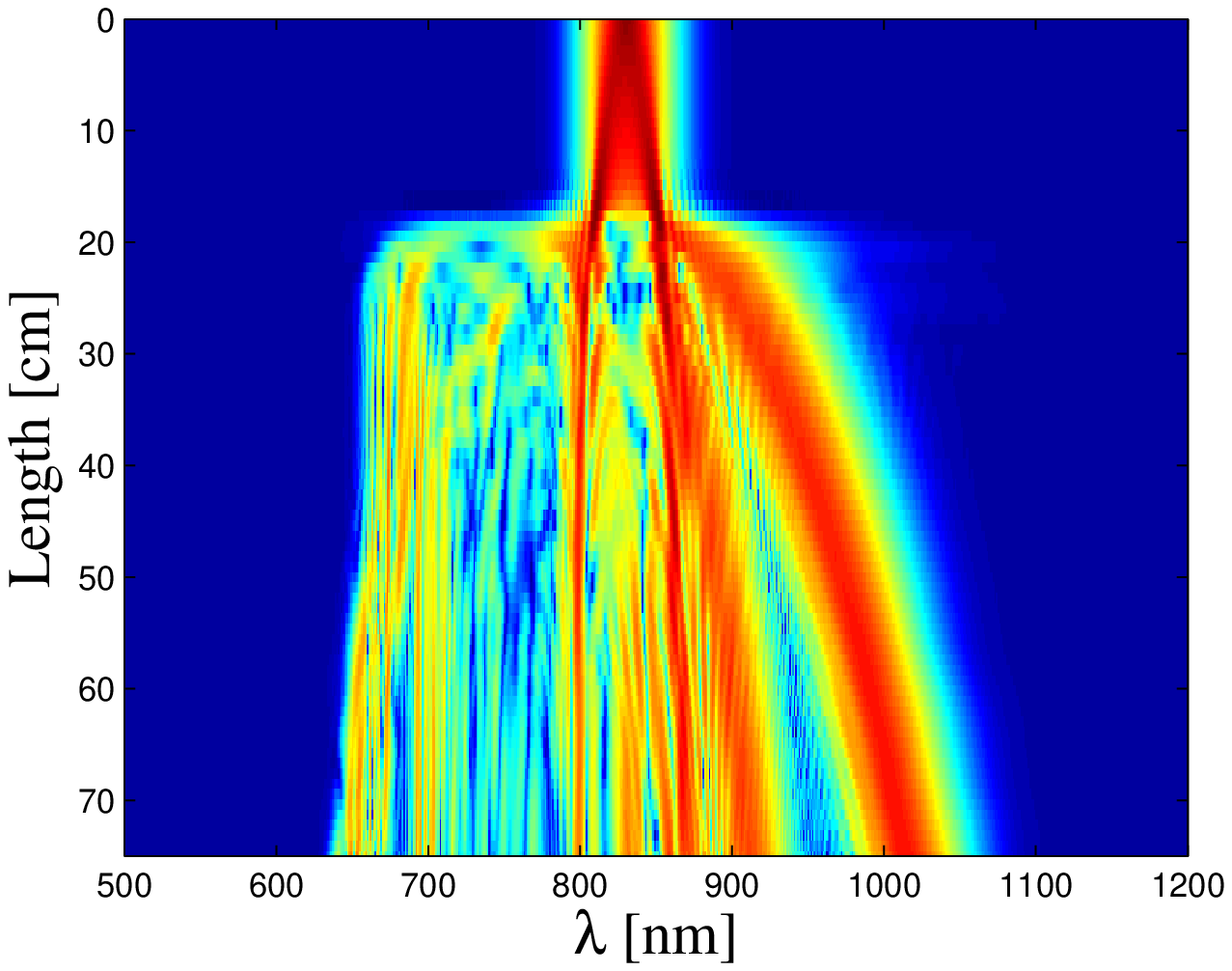}
  \caption{Simulated spectral evolution along the fiber length.
	Pump pulses with identical peak power produce more soliton shifting with unchirped (a) than with heavily chirped ($\sim$7400~fs$^2$) (b) pump pulses.
	}
  \label{fg:l_sim}
\end{figure}

In order to further analyze soliton formation, we studied the pulse evolution along the fiber by performing numerical simulations.
The spectrum evolution along the fiber, for a given coupled power, in the TL and the chirped cases are shown in Fig. \ref{fg:l_sim} (a) and (b), respectively. 
These simulations show that in the case of chirped pump pulses ($\sim7400$~fs$^2$), the spectrum broadening and the soliton formation take place farther down into the fiber (see Fig. \ref{fg:l_sim}), as compared to the TL case. 

The delay in the formation of the soliton can be explained by an interplay of opposite chirping effects: the positive chirp acquired by traversing the AOM is compensated as the pulse advances into the PCF, in anomalous propagation, leading to pulse compression.
The PCF itself provides pulse compression in the first stretch of the fiber previously to the branching of a soliton. 
Therefore, the SSFS effective length, i.e., the fiber path where nonlinearity broadens the spectrum, is longer in the TL case.
If the chirp is overcompensated and a negatively chirped pulse is fed into the fiber, these pulses would also be compressed within the first stretch of the fiber due to SPM \cite{Washburn:00} leading to the same behavior than in the positively chirped case, resulting in a narrower tunability range.

Once the soliton is formed and the peak power is high enough, intrapulse Raman scattering red-shifts the soliton as it propagates through the remaining of the fiber.
This spectral shift increases with both fiber length and soliton peak power \cite{Agrawal}. 
So the fact that the soliton is formed at different lengths explains the red shifts observed for the same coupled power.

However, as a larger wavelength shift can be achieved with a higher input power, this shortening in the effective length in the chirped case could be compensated by coupling more power into the PCF \cite{Nishizawa1999}.
Another possibility for compensating this effect on the SSFS is using a longer PCF. 

\subsection{Fiber power conversion efficiency} \label{sec:conv}

\begin{figure}[htbp]
  \centering
  \includegraphics[width=0.45\textwidth]{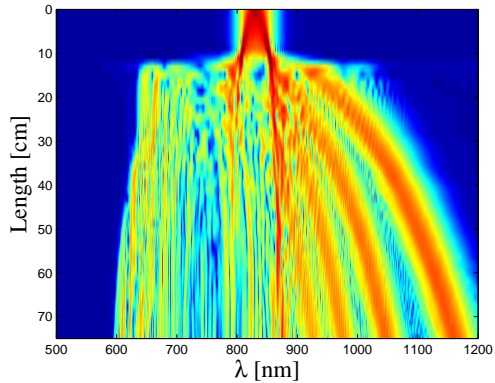}
  \caption{Simulation of the spectral evolution along the fiber showing that the same shifting shown in Fig. \ref{fg:l_sim} (a) for unchirped pump pulses is attained with chirped ones ($\sim7400$~fs$^2$) with a higher pump power.}
  \label{fg:cua}
\end{figure}
 
Figure \ref{fg:cua} shows a simulation where the same shift wavelength obtained for TL pump pulses is achieved by increasing the coupled power in the shorter effective length fiber ($7400$~fs$^2$ chirp).
Fission of more than one soliton branch is visible in this case, as compared to the case of TL pump pulses, for which only a single soliton branch appears (Fig. \ref{fg:l_sim}). 
Each soliton branch carries a fundamental soliton ($N=1$) with a peak power $P_0$ given by \cite{Agrawal}
\begin{equation}
  N^2=1=\frac{\gamma P_0 T_0^2}{|\beta_2|},
\end{equation}
As $\gamma$, $\beta_2$ and $T_0$ (see Figs. \ref{fg:dt_vs_lo_TL} and \ref{fg:chirp_dt_dl}) are the same in both, the TL and chirped cases, the peak power of the solitons is also the same.

The arising of new soliton branches partially accounts for the increased pump power required in the chirped case (Fig. \ref{fg:cua}) to attain the same shift.
Indeed, the soliton-pump power ratio is $0.2$ in the chirped case and $0.44$ in the TL case.
This result reveals that the use of the PCF as a compressor decreases its power conversion efficiency.

On the other hand, it is possible to achieve the same soliton shift as in the TL case by increasing the fiber length, and keeping the same pump power.
In this case, the power conversion efficiency is even lower, $0.17$, as predicted by simulations.

\section{Conclusions}\label{sec:conc}

We have presented a high-speed tunable soliton infrared source capable of generating $\sim45$~fs transform-limited pulses in the range from $860$ to $1160$~nm.
Both the pulsewidth and the spectral width were shown to remain constant over the entire tuning range, even when pump pulses were heavily chirped up to $7400$~fs$^2$. 
Insensitivity to the chirp of pump pulses points out to the feasibility of avoiding bulky compensation optics prior to the PCF, opening up the possibility to build reliable and compact high-speed tunable femtosecond sources in the near infrared region.
A minor drawback of this source is that either more power needs to be coupled or a longer PCF needs to be used in order to achieve the same tuning range obtained with transform-limited pump pulses.

\begin{acknowledgements}
This work was supported by ANPCyT PICT 2006-1594, ANPCyT PICT 2006-497 and UBA Programaci\'on Cient\'\i{}fica 2008-2010, Proyecto N X022.
\end{acknowledgements}


\begin{thebibliography}{10}
\newcommand{\enquote}[1]{\textit{#1}}

\bibitem{Nishizawa1999}
N Nishizawa, T Goto, \enquote{Compact system of wavelength-tunable femtosecond soliton pulse generation using optical fibers,} IEEE Photon.
  Technol. Lett. \textbf{11}, 325 (1999).

\bibitem{Abedin2004a}
K Abedin, F Kubota, \enquote{Wavelength tunable high-repetition-rate picosecond and femtosecond pulse sources based on highly nonlinear photonic crystal fiber,} IEEE J. Sel. Topics Quantum Electron. \textbf{10}, 1203 (2004).

\bibitem{Lee2008}
J H Lee, J van Howe, C Xu, X. Liu, \enquote{Soliton self-frequency shift: Experimental demonstrations and applications,} IEEE J. Sel. Topics Quantum Electron. \textbf{14}, 713 (2008).

\bibitem{Agrawal}
G P Agrawal, \emph{Nonlinear fiber optics}, Academic Press, San Diego (2007).

\bibitem{Mitschke:86}
F M Mitschke, L F Mollenauer, \enquote{Discovery of the soliton self-frequency shift,} Opt. Lett. \textbf{11}, 659 (1986).

\bibitem{soliton-teo-Gordon}
J P Gordon, \enquote{Theory of the soliton self-frequency shift,} Opt. Lett. \textbf{11}, 662 (1986).

\bibitem{Washburn2001}
B Washburn, S Ralph, P Lacourt, J Dudley, \enquote{Tunable near-infrared femtosecond soliton generation in photonic crystal fibers,}  Electronics Lett. \textbf{37}, 1510 (2001).

\bibitem{Takayanagi:06}
J Takayanagi, T Sugiura, M Yoshida, N Nishizawa, \enquote{1.0-1.7 $\mu$m wavelength-tunable ultrashort-pulse generation using femtosecond yb-doped fiber laser and photonic crystal fiber,} IEEE Photon. Technol. Lett. \textbf{18}, 659 (2006).

\bibitem{Russell03}
P Russell, \enquote{Photonic crystal fibers,} Science \textbf{299}, 358 (2003).

\bibitem{Skryabin03}
D V Skryabin, F Luan, J C Knight, P St J Russell, \enquote{Soliton self-frequency shift cancellation in photonic crystal fibers,} Science \textbf{31}, 1705 (2003).

\bibitem{Nishi02}
N Nishizawa, Y Ito, T Goto, \enquote{0.78-0.90 wavelength-tunable femtosecond soliton pulse generation using photonic crystal fiber,} IEEE Photon. Technol. Lett. \textbf{14}, 986 (2002).

\bibitem{Abedin:03}
K S Abedin, F Kubota, \enquote{Widely tunable femtosecond soliton pulse generation at a 10-ghz repetition rate by use of the soliton self-frequency shift in photonic crystal fiber,} Opt. Lett. \textbf{28}, 1760 (2003).

\bibitem{Ishii2006}
N Ishii, C Y Teisset, E E Serebryannikov, T Fuji, T Metzger, F Krausz, A M Zheltikov, \enquote{Widely tunable soliton frequency shifting of few-cycle laser pulses,} Phys. Rev. E \textbf{74}, 036617 (2006).

\bibitem{fs_soliton_source}
M E Masip, A A Rieznik, P G K{\"o}nig, D F Grosz, A V Bragas, O E Mart{\'i}nez, \enquote{Femtosecond soliton source with fast and broad spectral tunability,} Opt. Lett. \textbf{34}, 842 (2009).

\bibitem{Sanders2002}
S Sanders, \enquote{Wavelength-agile fiber laser using group-velocity dispersion of pulsed super-continua and application to broadband absorption spectroscopy,} Appl. Phys. B Lasers Opt. \textbf{75}, 799 (2002).

\bibitem{Walewski2004}
J Walewski, M Borden, S Sanders, \enquote{Wavelength-agile laser system based on soliton self-shift and its application for broadband spectroscopy,} Appl. Phys. B Lasers Opt. \textbf{79}, 937 (2004).

\bibitem{Sumimura:08}
K Sumimura, T Ohta, N Nishizawa, \enquote{Quasi-super-continuum generation using ultrahigh-speed wavelength-tunable soliton pulses,} Opt. Lett.\textbf{33}, 2892 (2008).

\bibitem{Sumimura2010}
K Sumimura, Y Genda, T Ohta, K Itoh, N Nishizawa, \enquote{Quasi-supercontinuum generation using 1.06 $\mu$m ultrashort-pulse laser system for ultrahigh-resolution optical-coherence tomography.} Opt. Lett. \textbf{35}, 3631 (2010).

\raggedbottom
\pagebreak

\bibitem{chan:08}
M-C Chan, S-H Chia, T-M Liu, T-H Tsai, M-C Ho, A Ivanov, A Zheltikov, J-Y Liu, H-L Liu, C-K Sun, \enquote{1.2- to 2.2- m tunable raman soliton source based on a cr:forsterite laser and a photonic-crystal fiber,} IEEE Photon. Technol. Lett. \textbf{20}, 900 (2008).

\bibitem{Nicholson:04}
J Nicholson, A Yablon, P Westbrook, K Feder, M Yan, \enquote{High power, single mode, all-fiber source of femtosecond pulses at 1550 nm and its use in supercontinuum generation,} Opt. Express \textbf{12}, 3025 (2004).

\bibitem{prisma_OEM}
R L Fork, O E Mart{\'i}nez, J P Gordon, \enquote{Negative dispersion using pairs of prisms,} Opt. Lett. \textbf{9}, 150 (1984).

\bibitem{brito-cruz}
R L Fork, C H B Cruz, P C Becker, C V Shank, \enquote{Compression of optical pulses to six femtoseconds by using cubic phase compensation,} Opt. Lett. \textbf{12}, 483 (1987).

\bibitem{Chilla:91}
J L A Chilla, O E Martinez, \enquote{Direct determination of the amplitude and the phase of femtosecond light pulses,} Opt. Lett. \textbf{16}, 39 (1991).

\bibitem{Costantino_fs}
S Costantino, A R Libertun, P D Campo, J R Torga, O E Mart{\'i}nez, \enquote{Fast scanner with position monitor for large optical delays,} Opt. Comm. \textbf{198}, 287 (2001).

\bibitem{Dudley_06}
J Dudley, G Genty, S Coen, \enquote{Supercontinuum generation in photonic crystal fibers,} Rev. Mod. Phys. \textbf{78}, 1135 (2006).

\bibitem{Heidt:09}
A Heidt, \enquote{Efficient adaptive step size method for the simulation of supercontinuum generation in optical fibers,} J. Lightwave Technol. \textbf{27}, 3984 (2009).

\bibitem{Lin_06}
Q Lin, G Agrawal, \enquote{Raman response function for silica fibers,} Opt. Lett. \textbf{31}, 3086 (2006).

\bibitem{Washburn:00}
B R Washburn, J A Buck, S E Ralph, \enquote{Transform-limited spectral compression due to self-phase modulation in fibers,} Opt. Lett. \textbf{25}, 445 (2000).


\end{thebibliography}

\end{document}